\documentclass{INTERSPEECH2023}
\usepackage[linesnumbered,ruled]{algorithm2e}
\usepackage{multirow}
\usepackage{array}
\pdfoutput=1

\interspeechcameraready

\title{Interpretable Style Transfer for Text-to-Speech with ControlVAE and Diffusion Bridge}
\name{Wenhao Guan$^1$, Tao Li$^1$, Yishuang Li$^2$, Hukai Huang$^1$, Qingyang Hong$^{1*}$, Lin Li$^{2,3*}$ \thanks{*Corresponding authors}}
\address{
  $^1$School of Informatics, Xiamen University, China\\
  $^2$Institute of Artificial Intelligence, Xiamen University, China \\
  $^3$School of Electronic Science and Engineering, Xiamen University, China}
\email{\{qyhong,lilin\}@xmu.edu.cn}
\begin{document}

\maketitle
 
\begin{abstract}
With the demand for autonomous control and personalized speech generation, the style control and transfer in Text-to-Speech (TTS) is becoming more and more important. 
In this paper, we propose a new TTS system that can perform style transfer with interpretability and high fidelity.
Firstly, we design a TTS system that combines variational autoencoder (VAE) and diffusion refiner to get refined mel-spectrograms. Specifically, a two-stage and a one-stage system are designed respectively, to improve the audio quality and the performance of style transfer.
Secondly, a diffusion bridge of quantized VAE is designed to efficiently learn complex discrete style representations and improve the performance of style transfer.
To have a better ability of style transfer, we introduce ControlVAE to improve the reconstruction quality and have good interpretability simultaneously.
Experiments on LibriTTS dataset demonstrate that our method is more effective than baseline models \footnote{Audio samples are publicly available at \url{https://gwh22.github.io/}.}. 
\end{abstract}
\noindent\textbf{Index Terms}: speech synthesis, style transfer, variational autoencoder, diffusion probabilistic model

\section{Introduction}
With the rapid development of deep learning, the common speech synthesis task that only takes text as input
has recently almost reached the human level in the public dataset \cite{tan2022naturalspeech}. More researchers focused on the controllable and expressive TTS \cite{zhang2019learning,an2022disentangling} because the controllability provides users with more flexibility, and the improvement of expressiveness makes people have better auditory experience.

At present, most TTS systems \cite{ren2019fastspeech,renfastspeech,kim2020glow,kim2021conditional} can model controllable style attributes in different ways. FastSpeech \cite{ren2019fastspeech} which is a non-autoregressive model uses the trained autoregressive model as the teacher model to train the duration predictor, and utilizes the length regulator to control the duration information, thus indirectly controls the overall style. 
FastSpeech2 \cite{renfastspeech} employs the external duration aligner Montreal Forced Aligner (MFA) \cite{mcauliffe2017montreal} to train the duration predictor on the basis of FastSpeech, furthermore, it trains the pitch predictor, energy predictor and predictors of other styles in a supervised way to achieve more precise style control. 
Most TTS system pipelines have two modules. The first module is to produce intermediate representations and the second module is to generate raw waveforms conditioned on the intermediate representations.
VITS \cite{kim2021conditional} connects the two modules, generated by VAE \cite{kingma2013auto}, through latent variables to enable efficient end-to-end training, and it proposes a stochastic duration predictor to synthesize speech with different variations. 
However, they don't do style transfer tasks so they do not need reference speech as input. Subsequently, there were many works focused on designing powerful style encoders for style transfer tasks. Some works based on autoregressive models obtain fine-grained style representations by designing multi-level or multi-scale style modeling methods \cite{sun2020generating,sun2020fully,li21r_interspeech}. Due to the problem of low decoding speed in the autoregressive models, the following style modeling tasks mostly adopt a non-autoregressive architecture\cite{ren2019fastspeech,renfastspeech,jeong2021diff}.    Meta-StyleSpeech \cite{min2021meta} adopts the base architecture upon FastSpeech2, applying style adaptive layer norm and meta-learning algorithm to effectively synthesize style-transferred speech. Styler \cite{lee2021styler} models style factors by speech decomposition via information bottleneck.  GenerSpeech \cite{huang2022generspeech} proposes a multi-level style adaptor and a generalizable content adaptor to efficiently model the style information. Norespeech \cite{yang2022norespeech} proposes a knowledge distillation based conditional diffusion model to generate style representation from noisy reference audios.  But these methods often cannot perform style interpretability.
Some works \cite{zhang2019learning,wang2018style} are proposed to demonstrate that they have the ability of style interpretability and transfer simultaneously. Global style token (GST) \cite{wang2018style} designs a style token layer and a  reference encoder to explore the expressiveness of TTS systems unsupervisedly. VAE-Tacotron \cite{zhang2019learning} learns the style representation through VAE \cite{kingma2013auto}. A recent work Style-Label-Free \cite{10038135} extends VAE-Tacotron by proposing Quantized VAE and speaker-wise normalization in cross-speaker style transfer.
In this paper, we focus on high-fidelity style transfer and interpretability in speech synthesis which requires a better style representation and a interpretable disentangled style latent space. We utilize VAE based style encoder to have  access to the interpretable latent space and integrate VAE within diffusion probabilistic models (DPM) \cite{ho2020denoising} to overcome the over-smoothness problem  \cite{ren2022revisiting}. We also propose a diffusion bridge of Quantized VAE to improve the diversity of generated style representations.
We further introduce ControlVAE \cite{shao2020controlvae} to our system instead of original VAE to have better  reconstruction quality and good interpretability.

Our main contributions can be summarized as follows:
\begin{itemize}
\item We propose a new TTS system that incorporates VAE and DPM to get refined mel-spectrograms. Specifically, a two-stage and a one-stage training pipeline are designed respectively to improve the performance of style transfer.
\item A diffusion bridge of Quantized VAE is proposed to  model the diversity of style representations in latent space so that the TTS system achieves better performance of style transfer.
\item We introduce ControlVAE in our system to replace original VAE to a better reconstruction ability so that  improve the style transfer quality and have good style interpretability.
\end{itemize}

\section{Related Work}

\subsection{Diffusion probabilistic models}
The concept of diffusion was first defined in \cite{sohl2015deep} and then researchers proposed denoising diffusion probabilistic model (DDPM) \cite{ho2020denoising} which greatly promotes the development of generative models. 
DDPM has recently succeeded to advance the state-of-the-art results in speech synthesis \cite{jeong2021diff,huang2022prodiff}.

Diffusion process and reverse process are given by diffusion probabilistic models, which could be used for the denoising neural networks $\theta$ to learn data distribution. 
Similar as previous work \cite{sohl2015deep,ho2020denoising}, we define the data distribution as $q(x_{0})$.
Let $x_{1},\cdots,x_{T}$ be a sequence of variables with the same dimension.
The diffusion process is defined by a fixed Markov chain from data $x_{0}$ to the latent variable $x_{T}$:
\begin{equation}
q(x_{t}|x_{t-1})=N(x_{t};\sqrt{1-\beta_{t}}x_{t-1},\beta_{t}I)
\end{equation}
\begin{equation}
q(x_{1},\cdots,x_{T}|x_{0})=\prod_{t=1}^{T}q(x_{t}|x_{t-1})
\end{equation}
The reverse process aims to recover samples from Gaussian noise, which is Markov chain from $x_{T}$ to $x_{0}$ parameterized by shared $\theta$:
\begin{equation}
p_{\theta}(x_{t-1}|x_{t})=N(x_{t-1};\mu_{\theta}(x_{t},t),\sigma_{t}^2I)
\end{equation}
\begin{equation}
p_{\theta}(x_{0},\cdots,x_{T-1}|x_{T})=\prod_{t=1}^{T}p_{\theta}(x_{t-1}|x_{t})
\end{equation}
where $\alpha_{t}=1-\beta_{t}, \bar{\alpha}_{t}=\prod_{t=1}^{t}\alpha_{t}$, $\mu_{\theta}$ and $\sigma_{t}^2$ represent  the mean and standard derivation respectively. 
Finally, we can get the training objective as follows:
\begin{equation}
L_{DDPM}=E_{t,x_{0},\epsilon}[||\epsilon-\epsilon_{\theta}(\sqrt{\bar{\alpha}_{t}}x_{0}+\sqrt{1-\bar{\alpha}_{t}}\epsilon,t)||_{2}^{2}]
\end{equation}
where $\epsilon$ is the Gaussian noise and $\epsilon_{\theta}(\cdot)$ is the output of model.
For sampling phase, the sampling formulation is computed as follows:
\begin{equation}
x_{t-1} = \frac{1}{\sqrt{\alpha_{t}}}(x_{t}-\frac{\beta_{t}}{\sqrt{1-\bar{\alpha}_{t}}}\epsilon_{\theta}(x_{t},t))+\sigma_{t}z,
\end{equation}
where $\epsilon \sim N(0,I)$, $p_{z}=N(z;0,I)$ and $\sigma_{t}=\sqrt{\frac{1-\bar{\alpha}_{t-1}}{1-\bar{\alpha_{t}}}\beta_{t}}$. As a result, the final data distribution $p_{x_{0}}$ is obtained through iterative sampling over all of the time steps.

\subsection{Variational Autoencoder}
VAE has been applied for latent representation learning of natural speech for years \cite{zhang2019learning,10038135}.
It is assumed that the observed data distribution $p(x)$ is generated by some random process from a random latent variable $z$. The true posterior distribution $p_{\theta}(z|x)$ is intractable because of the undifferentiable marginal likelihood $p_{\theta}(x)$. To address this problem, $q_{\phi}(z|x)$ is introduced as an approximation to the true posterior distribution $p_{\theta}(z|x)$. Finally, we can get the formulation of $logp_{\theta}(x)$:
\begin{align} \label{H}
\begin{split}
logp_{\theta}(x) & \geq E_{q_{\phi}(z|x)}[log\frac{p_{\theta}(x,z)}{q_{\phi}(z|x)}]\\
 &= E_{q_{\phi}(z|x)}[logp_{\theta}(x|z)]-D_{KL}(q_{\phi}(z|x)||p_{\theta}(z))
\end{split}
\end{align}

The encoder of the VAE is chosen to model a multivariate Gaussian with diagonal covariance, and the prior is often assumed to be a standard multivariate Gaussian:
\begin{equation}
q_{\phi}(z|x)=N(z;\mu_{\phi}(x),\sigma_{\phi}^{2}(x)I)
\end{equation}
\begin{equation}
p_{z}=N(z;0,I)
\end{equation}
where $\mu(x)$ and $\sigma^{2}(x)$ in $q_{\phi}(z|x)$ are learned via neural network, and the reparameterization trick is introduced to VAE framework to avoid non-derivable problem. Thus, each $z$ is computed as a deterministic function of input $x$ and auxiliary noise variable $\epsilon$, where $\odot$ represents an element-wise product.
\begin{equation}
z=\mu_{\phi}(x)+\sigma_{\phi}(x)\odot\epsilon
\end{equation}


\subsection{Quantized VAE}
Quantized VAE is proposed in \cite{10038135} to improve the representation ability of original VAE encoder.
Quantized VAE simply extends VAE  by adding a discrete codebook component to the VAE output $z$. $z$ is compared with all the vectors in the codebook, and the closest codebook vector is fed into the VAE decoder. The vector quanatization (VQ) loss which consists of commitment loss and codebook loss is as follows:
\begin{equation}
L_{Q}=||sg[z]-q||_{2}^{2}+\gamma||z-sg[q]||_{2}^{2},
\end{equation}
where $z$ is referred as VAE output, $q$ is referred as the codebook vector, $\gamma$ represents the weight of commitment loss, $sg[\cdot]$ refers to the operation of stop gradient.

\section{Proposed Method}
\begin{figure*}[htbp]
\centering
\includegraphics[scale=0.411]{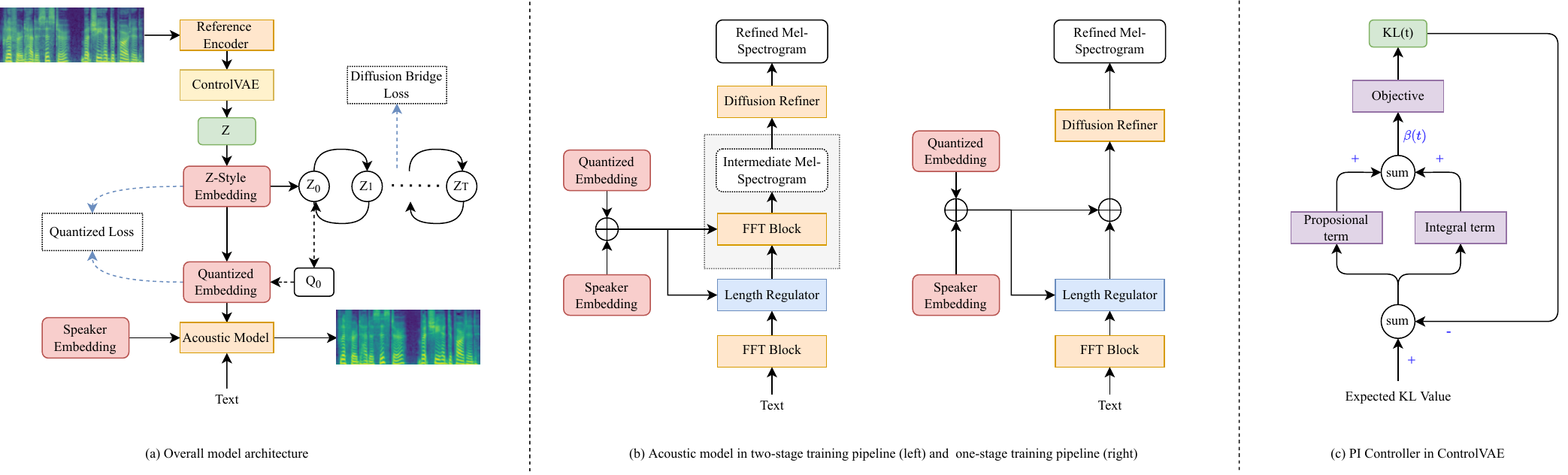} 
\caption{(a) the overall architecture of IST-TTS. (b)  the acoustic model in two-stage training pipeline (left) and one-stage training pipeline (right). (c)  the PI controller algorithm in ControlVAE.}
\label{fig1}
\end{figure*}


Figure 1 shows our proposed TTS method named IST-TTS (The abbreviation of Interpretable Style Transfer for Text-to-Speech). Our proposed method consists of three parts: the (1) diffusion refiner, (2) diffusion bridge, and (3) ControlVAE.
\subsection{Model architecture}
The overall architecture of our method is illustrated in Figure \ref{fig1} (a). A reference Mel-spectrogram is fed into a reference encoder to extract style information,  and the style information is then pass through ControlVAE to obtain the interpretable latent space $Z$. The quantized embedding is obtained by proposed diffusion bridge, and it is fed into the acoustic model. We use the same  architecture for diffusion bridge as \cite{kongdiffwave} to learn diverse style representations. 

The acoustic model is shown as Figure \ref{fig1} (b), which is based on the architecture of FastSpeech \cite{ren2019fastspeech}. And a diffusion refiner \cite{jeong2021diff} is designed for two-stage and one-stage training pipeline to explore the effectiveness of combining VAE and DPM.
Additionally, the speaker embedding is extracted by x-vector \cite{snyder2018x}. 
Note that we use MFA \cite{mcauliffe2017montreal} to replace the previous distillation method to effectively train the duration predictor.

\subsection{Diffusion Refiner}
We propose to incorporate VAE framework and DPM in two-stage and one-stage training pipeline respectively to explore the effectiveness of our method.

\subsubsection{Two-stage and one-stage training pipeline}
As shown in the left of Figure \ref{fig1} (b), in two-stage training pipeline, the model firstly generates an intermediate mel-spectrogram and it can be fed into the vocoder to get the intermediate waveform, the model is named as  VAEFS in this paper. And then the intermediate mel-spectrogram which is processed by a linear layer is fed into diffusion refiner as the condition of diffusion model, the model is named as VAEFS+2s in this paper.

As shown in the right of Figure \ref{fig1} (b), in one-stage training pipeline, it will degrade into the method of Diff-TTS \cite{jeong2021diff} in acoustic model, which is named as VAEFS+1s in this paper. Note that the designs in the rest of this paper are based on the acoustic model in one-stage training pipeline.

\subsubsection{Conditional diffusion model}
Diffusion refiner in our method is a conditional diffusion model because the input needs external intermediate mel-spectrogram or decoder input to be as the condition of diffusion model. $c$ represents the condition. The training objective is as follows:
\begin{equation}
L_{R}=E_{t,x_{0},\epsilon,c}[||\epsilon-\epsilon_{\theta}(\sqrt{\bar{\alpha}_{t}}x_{0}+\sqrt{1-\bar{\alpha}_{t}}\epsilon,t,c)||_{2}^{2}]
\end{equation}

\subsection{Diffusion Bridge}
Quantized VAE discretizes the latent features using vector quantization to generate more expressive samples. Inspired by recent work \cite{cohen2022diffusion}, we propose a new Diffusion Bridge to improve the ability of expressiveness in Quantized VAE. Specifically, we consider using diffusion models in a continuous space, which is the $z$ latent space of VAE output in our work, to efficiently learn complex discrete distributions. Note that the sampling process of diffusion bridge is only used in inference stage. The process is shown in Figure \ref{fig1} (a) and the training loss is:
\begin{equation}
L_{B}=E_{t,z_{0},\epsilon}[||\epsilon-\epsilon_{\theta}(\sqrt{\bar{\alpha}_{t}}z_{0}+\sqrt{1-\bar{\alpha}_{t}}\epsilon,t)||_{2}^{2}]
\end{equation}

\subsection{ControlVAE}


The original VAE models may suffer from KL vanishing and low reconstruction quality problem. To address the issuses, we propose ControlVAE \cite{shao2020controlvae} which combines a controller with the basic VAE as an alternative of original VAE. 
 Specifically, a new non-linear proportional-integral (PI) controller 
is designed to automatically tune the weight added in the VAE objective using the output KL-divergence as feedback during model training. The PI controller algorithm can be illustrated as Figure \ref{fig1} (c), the weight $\beta(t)$ and the loss of ControlVAE are as follows:
\begin{equation} \label{beta}
\beta(t)=\frac{K_{p}}{1+exp(e(t))}-K_{i}\sum_{j=0}^{t}e(j)+\beta_{min}
\end{equation}
where $K_{p}$ and $K_{i}$ represent the coefficient of propositional term and integral term. $e(t)$ is the error between current KL value and expected KL value. $\beta_{min}$ is a constant.
\begin{equation} \label{C}
L_{C}=E_{q_{\phi}(z|x)}[logp_{\theta}(x|z)]-\beta(t)D_{KL}(q_{\phi}(z|x)||p(z))
\end{equation}

The reconstruction loss is computed with the help of auxiliary Feed-Forward Transformer Decoder like DiffSpeech\cite{liu2022diffsinger} in one-stage training pipeline.

Finally, the total training loss of our method is:
\begin{equation}
L_{All}=L_{C}+L_{R}+L_{Q}+L_{B}
\end{equation}

\begin{table}[!t]
	  \centering  
	  \caption{The objective results of style transfer. The last three rows show the results of ablation study.}
	  \label{t1} 
	  \begin{center}
	   \resizebox{46mm}{12mm}{
		\begin{tabular}{ccc}\hline 
 		Method & FD $\downarrow$ & MCD $\downarrow$  \\ \hline
 		VAEFS & 6.36 & 5.82 \\
 		VAEFS+2s & 3.69 & 5.79 \\
 		VAEFS+1s & 4.47 & 5.84 \\ \hline
 		GenerSpeech & 4.83 & 5.66 \\ 
 		IST-TTS (ours) & \textbf{3.37} & \textbf{5.63} \\ \hline
 		w/o ControlVAE & 3.60 & 5.90 \\
 		w/o VQ & 4.27 &  6.21 \\
 		w/o Diffusion Bridge & 3.81 & 5.67 \\
 		\hline
		\end{tabular}}
	\end{center}
\end{table}

\begin{table*}[!ht]
	  \centering  
	  \caption{The subjective results of style transfer. 1s and 2s represent one stage and two stage training with diffusion refiner respectively. B,S,O denote baseline, same and ours respectively. The last three rows show the results of ablation study.}
	  \label{t2} 
	  \begin{center}
	  \resizebox{127mm}{17mm}
	  {
		\begin{tabular}{c|cc|c|ccc|c|ccc}\hline \multicolumn{1}{c|}{\multirow{3}{*}{Method}} & \multicolumn{2}{c|}{5-scale Score} & \multicolumn{4}{c|}{Parallel Style Transfer} & \multicolumn{4}{c}{Non-Parallel Style Transfer}\\ 
\cline {2-11}
& MOS & SMOS & 7-point score & \multicolumn{3}{c|}{ Preference (\%)} & 7-point score & \multicolumn{3}{c}{ Preference (\%)} \\
& & & & B & S & O & & B & S & O \\ \hline
 		Reference & 4.52$\pm$0.062 & $-$ & \multicolumn{8}{c}{ $-$} \\
 		Reference(Mel+Voc) & 4.49$\pm$0.076 & 4.41$\pm$0.065 & \multicolumn{8}{c}{ $-$} \\
 		\hline
 		VAEFS & 3.31$\pm$0.081 & 2.85$\pm$0.083 & 1.75 & 12 & 14 & 74 & 1.60 & 22 & 28 & 50\\
 		VAEFS+2s & 3.75$\pm$0.092 & 3.21$\pm$0.076 & 1.30 & 15 & 18 & 67 & 1.52 & 26 & 33 & 41\\
 		VAEFS+1s & 3.79$\pm$0.079 & 3.26$\pm$0.105 & 1.21 & 20 & 25 & 55 & 1.32 & 15 & 38 & 47\\
 		\hline
 		GenerSpeech  & 3.96$\pm$0.086 & 3.99$\pm$0.089 & 0.21 & 32 & 34 & 34 & 0.51 & 32 & 32 & 36\\ \hline
 		IST-TTS (ours) & \textbf{4.17$\pm$0.055} & \textbf{4.02$\pm$0.091} & \multicolumn{8}{c}{ $-$} \\
 		\hline
 		w/o ControlVAE & 4.09$\pm$0.056 & 3.86$\pm$0.083 & 0.76 & 25 & 35 & 40 & 0.60 & 26 & 40 & 34\\
 		w/o VQ & 3.85$\pm$0.061 & 3.81$\pm$0.076 & 0.91 & 19 & 37 & 44 & 0.69 & 29 & 36 & 35\\
 		w/o Diffusion Bridge & 4.01$\pm$0.081 & 3.79$\pm$0.073 & 0.71 & 30 & 27 & 43 & 0.55 & 34 & 28 & 38\\
 		
 		\hline
		\end{tabular}}
	\end{center}
\end{table*}

\section{Experimental Setup}
\subsection{Dataset}
As for dataset, we use LibriTTS \cite{Zen2019}, which is a multispeaker English corpus containing 586 hours of speech clips with 2456 speakers, to evaluate our method. 
And we convert the sampling rate to 22050Hz and extract the spectrogram with the FFT size of 1024, hop size of 256, and window size of 1024. 

\subsection{Training setting}
The proposed model was trained for 320K iterations using Adam optimizer \cite{kingma2014adam} on a single NVIDIA TELSA V100 GPU. Additionally, we utilize a pretrained HiFi-GAN \cite{kong2020hifi} as the neural vocoder to convert mel-spectrogram to waveform. 

The hyperparameters of reference encoder is the same as \cite{zhang2019learning}, the dimension of $z$, the codebook size and $\gamma$ of vector quantization are set to 32, 1024 and 0.25, the hyperparameters $\beta_{min}$, $K_{p}$, $K_{i}$ and expected KL value in ControlVAE are set to 0, 0.01, 0.0001 and 3 respectively. Exponential moving averages (EMA) \cite{van2017neural} is used instead of codebook loss in Quantized VAE to get faster convergence speed. We use cost annealing method in original VAE. $T$ in diffusion refiner (bridge) is set to 1000.

\begin{figure}[htbp]
\centering
\includegraphics[scale=0.67]{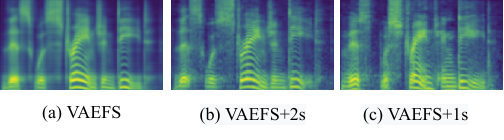} 
\caption{The comparison of (a) VAEFS, (b) VAEFS+2s and (c) VAEFS+1s.}
\label{fig2}
\end{figure}

\begin{figure}[htbp]
\centering
\includegraphics[scale=0.72]{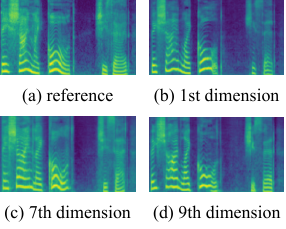} 
\caption{The generated mel-spectrograms for disentangling.}
\label{fig3}
\end{figure}

\subsection{Baseline models}
We compare the quality and similarity of generated audios by our IST-TTS with public baseline model GenerSpeech \cite{huang2022generspeech} and our proposed baseline models, VAEFS(+1s/2s).

\section{Experiment Results and Analysis} 

\subsection{Style Transfer Quality and Similarity}
We evaluate the style transfer quality and similarity by objective metrics and subjective evaluations. 
The objective metrics include frechet distance (FD) \cite{heusel2017gans} metric and mel cepstral distortion (MCD) metric. The FD in audio indicates the similarity between generated samples and target samples and MCD evaluates the compatibility between the spectra of two audio clips. As for subjective evaluations, we conduct 5-scale Mean Opinion Score (MOS) and Similarity Mean Opinion Score (SMOS) test between IST-TTS and the baselines. To further evaluate the style transfer performance, an AXY test used in \cite{huang2022generspeech} is conducted, the  range of 7-point score is from -3 to 3, 0 represents “Both are about the same distance".

\subsubsection{Parallel Style Transfer}
As for parallel style transfer, 
The objective results of FD and MCD shown in Table \ref{t1} show that our proposed IST-TTS outperform baseline models, indicating that the proposed model generates higher quality results.
The MOS,SMOS and parallel style transfer results in Table \ref{t2} show our IST-TTS achieve higher quality and similarity.

Furthermore, we show the results of comparison between VAEFS, VAEFS+2s and VAEFS+1s in Figure \ref{fig2}, inferring that our diffusion refiner can overcome the oversmoothness problem in VAEFS so that obtain better performance of quality.

\subsubsection{Non-parallel Style Transfer}
Non-parallel style transfer represents that the text is changed from the reference utterance.
The non-parallel style transfer results are shown in Table \ref{t2}, which denote that our method has a better performance in non-parallel style transfer than baseline models by using ControlVAE and diffusion bridge to get more expressive results. 
\subsubsection{Ablation Study}
As shown in the last three rows of Table \ref{t1} and Table \ref{t2}, we conduct ablation studies to demonstrate the effectiveness of several designs in IST-TTS, including ControlVAE, vector quantization (VQ) and diffusion bridge. Both the objective  metrics and subjective scores drop when removing VQ or diffusion bridge, and replacing ControlVAE with the original VAE results in decreased quality and similarity. These demonstrate the efficiency of the proposed method in modelling expressive style representations.

\subsection{Style Interpretability}
To evaluate the performance of style interpretability, we experiment on single dimension of ControlVAE latent space $z$ to explore the ability of disentangling as shown in Figure \ref{fig3} which represents some different speaking styles of energy, pitch variation and pitch level. 
The limitation in the analysis of style interpretability is that the dataset we chose does not have detailed style labels, so there is no way to conduct quantitative interpretable comparisons.
\section{Conclusions}
In this paper, we propose a TTS method for interpretable style transfer named IST-TTS, which incorporates VAE with diffusion refiner to improve the audio quality and the performance of style transfer. Furtherly we propose diffusion bridge to perform better style transfer. And finally we introduce ControlVAE to achieve better style transfer quality and good disentanglement. In the future, we will explore better method for interpretability.

\section{Acknowledgements}
Thanks to the National Natural Science Foundation of China (Grant No.62276220, No.62001405 and No.61876160) for funding.

\bibliographystyle{IEEEtran}
\bibliography{mybib}

\end{document}